\documentclass[%
reprint,
showkeys,showpacs,preprintnumbers,
amsmath,amssymb,
aps,
prb,
floatfix,
]{revtex4-1}

\usepackage{graphicx}
\usepackage{dcolumn}
\usepackage{bm}
\usepackage[dvips]{hyperref}
\usepackage[T1]{fontenc}
\usepackage[utf8]{inputenc}
\usepackage[active]{srcltx}
\begin{document}

\title{Towards the observation of phase locked Bloch oscillations in arrays of small Josephson junctions}

\author{Felix Maibaum}
 \email{felix.maibaum@ptb.de}
\author{Sergey V. Lotkhov}
\author{A. B. Zorin}%

\affiliation{%
 Phsyikalisch-Technische Bundesanstalt, Bundesallee 100, 38116 Braunschweig
}%

\date{\today}

\begin{abstract}
We have designed an experiment and performed extensive simulations and preliminary measurements to identify a set of realistic circuit parameters that should allow the  observation of constant-current steps  at $I=2ef$ in short arrays of small Josephson junctions under external AC drive of frequency $f$. Observation of these steps demonstrating phase lock of the Bloch oscillations with the external drive requires a high-impedance environment for the array, which is provided by on-chip resistors close to the junctions. We show that the width and shape of the steps crucially depend on the shape of the drive and the electron temperature in the resistors. 
\end{abstract}

\maketitle


\section{\label{sec:Introduction}Introduction}
Shortly after the discovery of the Josephson effect it had been understood that superconducting circuits including small Josephson junctions can demonstrate quantum behavior of the Josephson phase $\varphi$.\cite{Anderson64,Larkin84} During the past decade, this topic has been extensively investigated for quantum information applications, see, for example, Refs.~\onlinecite{makhlin01,Devoret2004}.
One of the remarkable manifestations of quantum behavior is a band energy spectrum of an isolated single Josephson junction.\cite{AZL,likharev84} This effect results from the periodic dependence of the Josephson potential energy $E_J \cos \varphi$ and finite kinetic energy $\hat{Q}^2/2C$ associated with the charge variable $\hat{Q} = -i(2e)\frac{\partial}{\partial \varphi}$ and junction capacitance $C$, and having the scale $E_c = e^2/2C$. The motion of a fictitious particle in this periodic potential is similar to that of an electron in a crystal lattice, with the charge and capacitance in the Josephson case playing the role of the momentum and mass of the particle, respectively.

Applying a constant current $\bar{I}$ (similar to a constant electric field for the crystal lattice analogy) should produce Bloch oscillations of voltage across the junction with frequency $f_B=\bar{I}/2e$. Each period of oscillations then corresponds to the transfer of one Cooper pair with net charge $2e$ through the junction. This phenomenon is dual to the AC Josephson effect in larger junctions with classical behavior, which is associated with the motion of single flux quanta $\Phi_0$ in the direction transverse to the supercurrent flow. Thus, similar to the phase locking of Josephson oscillations leading to Shapiro steps of constant voltage on the $IV$ curve of large junctions, \cite{Shapiro63} applying an alternating signal of frequency $f$ to a small (Bloch) junction should lead to the appearance of constant-current plateaus $I = 2emf$ with $m=0, \pm 1, \pm 2,\dots$.\cite{AZL,likharev84}

The major prerequisite for the experimental demonstration of Bloch oscillations in small Josephson junctions is achieving a sufficiently high impedance of the electromagnetic environment seen by the junction at characteristic frequencies, $|Z_e(\omega)| \gg R_Q$, where $R_Q=h/4e^2 \approx 6.45\,\rm{k}\Omega$ is the resistance quantum.\cite{Schmid83}
Unfortunately, the first attempts to experimentally demonstrate this effect by engineering a high-ohmic environment using on-chip resistive leads \cite{kuzmin91_2} only showed peculiarities in the derivative of the $IV$ curves when the AC signal was applied. Although these peculiarities were positioned at $2ef$, the observation of clear current steps was not possible. The main reason for this was believed to be the substantial thermal fluctuations in the resistors. Our results from section\,\ref{sec:sim_results} indeed show that this seems to be the main obstacle to designing an experiment which clearly demonstrates this effect. 

More recently, Nguyen et al.\,\cite{nguyen07} have succeeded in the demonstration of Bloch oscillations by injecting a displacement current $I_d$ into the island of a Bloch transistor \cite{Likharev86, Likharev87} through a capacitive gate. In this experiment, the linear ramp of voltage $V_g$, applied to the gate capacitance $C_g$ and yielding sufficiently high impedance $Z_e(\omega) = 1/i\omega C_g$, ensured the constant current $I_d = C_g \frac{dV_g}{dt}$ fed into the island. The readout of the Bloch oscillations in this circuit was possible at discrete points in time by applying a switching current technique.

Finding ways towards a clear observation of Bloch oscillations driven by a real DC source is the problem which we address in this paper. Our motivation for developing this concept is the better understanding of the dynamics of this macroscopic quantum system and improvement of the shape of the phase locking steps with the goal of their possible application for the fundamental standard of current operating on coherent tunneling of single Cooper pairs.

\section{\label{sec:Background}Background}
The physics of Bloch oscillations in small Josephson junctions is most transparent in the representation of quasicharge $q$ which plays a role similar to the quasimomentum in solids. The eigenenergies of the Josephson junction $E_n(q)$, $n=0, 1, ...$, are periodic functions of $q$ with  a period of $2e$, whereas the eigenstates $|q,n\rangle$ are the Bloch functions.\cite{AZL,likharev84} At sufficiently low temperature the system occupies the ground state $n=0$ and
the observable voltage across the junction
\begin{equation}
\label{Vq} V(q) = dE_0(q)/dq
\end{equation}
is an odd periodic function of $q$, whose shape depends on the ratio of characteristic energies,  $\lambda = E_J/E_c$. In the case of weak Josephson coupling, $\lambda \ll 1$, $V(q)$ has a sawtooth shape with a maximum amplitude $V_{c}$ approaching $e/C$ for $\lambda \rightarrow 0$. In the
strong Josephson coupling case, $\lambda \gg 1$, this function is approximated by the expression $V(q) = V_c \sin (\pi q/e)$ with a smaller amplitude $V_c \approx 2^{11/4}\pi^{1/2}\lambda^{3/4}\exp[-(8\lambda)^{1/2}](e/C) \ll e/C$. \cite{likharev84}

The dynamics of quasicharge had been analyzed earlier within the framework of an RSJ-like model derived for the zero-band ($n=0$) approximation and linear damping in Refs.~\onlinecite{AZL, likharev84}. The quasicharge $q$ in this model is equal to the total charge fed into the junction by a current source, i.e. $q=\int_0^t I(t') dt' + q_0$. For the equivalent serial circuit including a small Josephson junction, resistor, and DC and AC voltage sources, the equation of motion for quasicharge can be written as
\begin{equation}
\label{eq-motion-RSJ} R\dot{q} + V(q) = \bar{V} + V_{\textrm{ac}} + V_{\textrm{noise}},
\end{equation}
where dot means the time derivative and $R\dot{q}$ yields the voltage across the series resistance $R$. Assuming a sufficiently slow motion of $q$ strictly in the ground state, i.e. $h f_B/\Delta_\mathrm{min}  \ll 1$, where $\Delta_\mathrm{min}$ is the minimum energy gap between the first excited and the ground state $[E_1(q)-E_0(q)]$, which is $\approx E_J$ for $\lambda \lesssim 1$, we have omitted in this equation an inertia term $\propto \ddot{q}$ describing the effect of so-called Bloch inductance $L_B$.\cite{zorin2006}  Thus the characteristic frequency in this first-order differential equation describing the overdamped system  is determined solely by the rate of damping, i.e. $\omega_c = \pi V_c/eR$. This quantity can be also interpreted as a characteristic recharging rate for a non-linear Bloch capacitance which has the reverse value $C_B^{-1} = dV(q)/dq = d^2E_0/dq^2$. \cite{likharev84}

A serial bias resistance $R \gg R_Q$ also results in a noise term $V_{\textrm{noise}}$, causing a finite linewidth of the oscillations proportional to $1/R$. \cite{likharev84} Moreover, in the realistic case of rather large fluctuations,\cite{kuzmin91_2} the Coulomb blockade corners in the $IV$ curve at $V = \pm V_c$ become rounded and the shape of expected Shapiro-like steps, always having a size $\lesssim 1.2 V_c$,\cite{likharev-book} is deteriorated by the noise even more strongly than the blockade corners.

It is not easy, however, to fulfill in experiment the requirements of (i) high linear damping and (ii) a relatively low noise level, enabling the observation of clear current steps with flat central parts. First, the state-of-the-art fabrication technology for thin-film resistors allows reproducible manufacturing of resistive stripes about $w = 100$\,nm in width and with resistivity $\rho$ up to about 1\,k$\Omega/\Box$, yielding a specific resistance $r=\rho/w \approx 10$\,k$\Omega/\mu$m. A sufficiently high value of resistance $R = r \ell \gg R_Q$ requires a length $\ell$ of several tens of micrometers. With a specific stray capacitance to ground of about $c\approx60$\,aF/$\mu$m,\cite{kuzmin91, zorin2000} the accumulated capacitance quickly becomes much larger than the self-capacitance of the Josephson junction $C$, which is typically in the range $0.1-1.0$\,fF. Therefore, at characteristic frequencies $\omega_c$ of the process, the effect of stray capacitance is significant and the resistive stripe must be considered as an $RC$ transmission line leading to frequency-dependent damping. The equation of motion for such a circuit is
\begin{equation}
\label{eq-motion-RC-lineSJ} \int_0^\infty K(t')\dot{q}(t-t')dt' + V(q) = \bar{V} + V_{\textrm{ac}} + V_{\textrm{noise}},
\end{equation}
where the kernel $K(t)$ is a Fourier transform of the $RC$ line impedance $Z(\omega) = (r/j\omega c)^{1/2}\tanh[(j\omega c r )^{1/2}\ell]$.

Second, the requirement of relatively small noise can be met if the parameter $V_c$ is sufficiently large, i.e. $eV_c \gg k_BT^*$, where $T^*$ is the electron temperature of the resistor. Although for observation of the blockade part of the $IV$ curve the temperature $T^*$ may only slightly exceed the mixing chamber (MC) temperature of the dilution refrigerator $T_{\textrm{MC}}$, it increases dramatically due to Joule heating when the resistor carries a sufficiently large current of the order of 1\,nA. \cite{webster2008} The maximum size of the first ``Shapiro`` step given by the dual to the RSJ model with harmonic drive is about $V_c$ \cite{likharev-book} and the absolute maximum $\approx 1.16V_c$ is achieved for drive frequency $\omega \approx 2\omega_c$ and amplitude $I_\omega \approx 2GV_c$. \cite{Likharev71} The position of such a step corresponds to $\bar{I} = e\omega_c/\pi = GV_c$, which for typical parameters is of the order of 1\,nA.

\section{Small junction arrays}
With the present technology of fabrication, the natural way of improving visibility of the steps by increasing both characteristic energies $E_J$ and $E_c$ is difficult, because it simultaneously requires a smaller junction size and larger critical current. Whereas a decrease in size of our $\approx (100$nm$)^2$ junctions to about one quarter of this area would still be feasible, thereby quadrupling the charging energy, the required fourfold increase in Josephson energy $E_J$ would neccessitate a 16 times higher critical current density $J_c$, which becomes much harder to manufacture reliably and will result in much less uniform junctions. On the other hand, using a serial array of $N$ small Josephson junctions would increase effective $E_c$ without having to decrease the junction size, thereby enabling the independent adjustment of $E_J$ and $E_c$. This is only true as long as the array can be considered a single lumped element, so it is neccessary to examine under which conditions this is the case.

In sufficiently long arrays, the charge injected into the array takes the form of solitons carrying a charge of $\pm 2e$, the dynamics of which is described by a sine-Gordon equation.\cite{Averin91} With the junction capacitance $C$ and the stray capacitance of the metallic islands between the junctions to ground $C_0$, the size of such a Cooper-pair soliton can be expressed in the number of islands $\Lambda_s = (C/C_0)^{1/2}$ over which the charge $2e$ is mostly distributed.\footnote{Strictly speaking this is true for $\lambda \ll 1$, and becomes $\Lambda_s = (e/\pi V_c C_0)^{1/2}$, for $\lambda \approx 1$ and higher}$^,$\cite{Haviland96} Formation of solitons inside the array leads to multiple solutions of the corresponding sine-Gordon equation,\cite{Averin91} and thus to a multivalued voltage $V_a(q)$ across the array as a function of injected charge $q$. Moreover, this voltage saturates at a maximum value of about $\sqrt{(\pi e V_c/C_0)}$.\cite{Haviland96} An optimum drive of an array with such a multivalued $V(q)$ dependence which can ensure controlled motion of $2e$-solitons along the array would probably require several AC gates with well-determined mutual phase shifts.

\begin{figure}[t]
 \centering
 \includegraphics[width=\columnwidth]{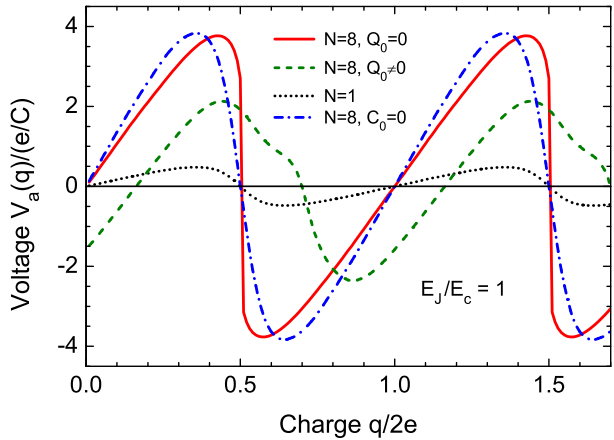}
 \caption{(Color online) The dependence $V_a(q)$ for a uniform array of $N=8$ junctions with $\lambda=1$ in the ideal case of negligible island capacitance (blue dash-dotted line), with island capacitance $C_0/C = 0.05$ calculated for zero offset charges (solid red line) and random distribution of these charges between -0.5$e$ and 0.5$e$ (dashed green line). For comparison, the dotted black line shows dependence $V(q)$ for a single junction. The curves were obtained by numeric solving of the corresponding sine-Gordon equation.\cite{Averin91}}
 \label{fig:Fig_V-vs-q}
\end{figure}

Still, one can use a relatively short array, $N < \Lambda_s$, which has a single-valued $2e$-periodic dependence $V_a(q)$ and, therefore, can be considered as a lumped element. 
 For zero offset charges on the inner islands the maximum value of $V_a(q)$ may approach almost the level $N$-times higher than that of the single junction, i.e. $NV_c$. \footnote{In practice, the maximum blockade voltage across the array $\hat{V}_a$ will only scale linearly up to $N \approx \Lambda_s/2$ and finally saturate when a full soliton fits inside the array. The practical limit for this will be around $N=10$ for typical parameters.} Figure\,\ref{fig:Fig_V-vs-q} 
shows several numerically obtained curves for $V_a(q)$, taking into account non-zero $C_0$ as well as one random configuration of offset charges.

\section{Preliminary measurements}\label{sec:measurements}
We fabricated Al/AlOx/Al junctions with an area of $\approx (100$nm$)^2$ defined by electron-beam lithography and angle evaporation. The junctions are arranged in short arrays of $N<10$ junctions. 
For these arrays, the capacitance to ground differs slightly between even- and odd-numbered islands, since one forms the top and one the bottom electrode for the junction, but it can be estimated at $\lesssim 14$aF per island on average on a 380$\mu$m thick silicon wafer,
  \footnote{Assuming a small wire with diameter $d$ over a ground plane at distance $h\gg d$, the capacitance per unit length is given by $C'\approx \frac{2\pi\epsilon_0\epsilon_r}{\ln(4h/d)}$, which yields 68.75\,aF/$\mu$m for $d=100$nm, $h=380\mu$m and $\epsilon_r=11.9$. This agrees well with the experimentally obtained 60aF/$\mu$m for the stray capacitance of the resistive microstrips.\cite{kuzmin91} Our junction arrays have similar dimensions, with one junction every 200nm, resulting in 13.75aF per island on average. This can be considered an upper bound, since this simple approximation neglects the fact that only one half space is filled with silicon. Approximations used for microstrip lines yield about half of this, but are generally not very accurate for  such large $h/d$.}
while the junction capacitances are around 0.5\,fF, yielding $\Lambda_s \approx 6$. The soliton size becomes even larger for moderately increased values of $E_J/E_c$, when the effect of nonlinear Bloch capacitance is taken into account.\cite{Haviland96, lotkhov07} When the chip is not glued directly onto grounded metal but instead on a PCB carrier, the stray capacitance is reduced further and the soliton size increases correspondingly. This means that we can probably use up to about eight junctions in series to increase effective $E_c$ of our arrays. 

\begin{figure}[t]
 \centering
 \includegraphics[width=\columnwidth]{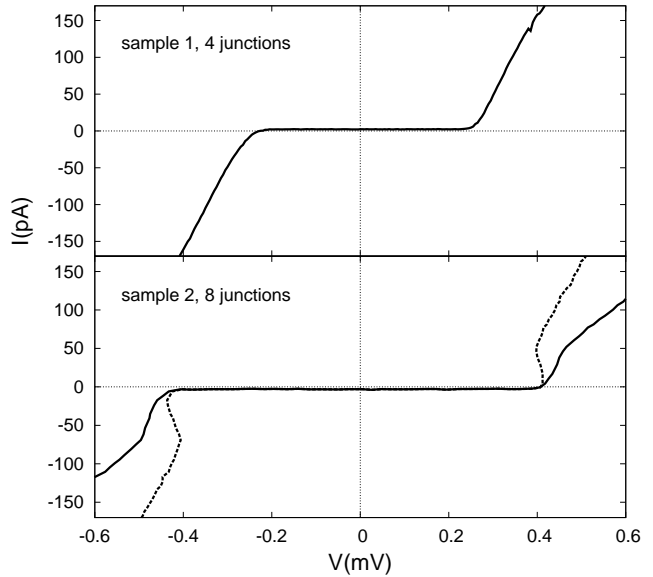}
 \caption{Measured $IV$ curves of a four junction array (top) and an eight junction array (bottom). The solid lines show the voltage across the array including bias resistors, the dashed line in the bottom panel shows the voltage across the array alone, forming a ``Bloch nose''. }
 \label{fig:bloch_nose}
\end{figure}

We measured several short arrays and observed scaling of the blockade voltage with the number of junctions in series, as well as clear backbending of the $IV$ curve, as shown in fig.\,\ref{fig:bloch_nose}.
The backbending indicates the presence of Bloch oscillations corresponding to the steady motion of a wave packet in the ground state, while the again increasing voltage of the array for currents above $\approx 50$\,pA shows that the upper Bloch bands become populated, presumably due to thermal excitation and Zener tunneling.\cite{Landau32,Zener32}
Thus we need to operate below this current for the zero-band model to be applicable. The relatively low current at which this inflection point occurs can be ascribed to the rather small energy gap ($<E_J$) between the ground state and the first excited state in an array with finite capacitance to ground.\cite{Geigenmuller89}

 The resistance of the array slightly above $T_c$ was $\approx~200$\,k$\Omega$. Using the Ambegaokar-Baratoff relation \cite{ambegaokar63} we calculated the Josephson energy of an individual junction to be about 35$\mu$eV, resulting in $\lambda\approx 0.8$ for an individual junction and an effective $\lambda_a\approx 0.1$ for the entire array when considered as a single equivalent junction.


Another important parameter is the electron temperature $T_e$ in the biasing resistors which provide the high-impedance environment. Due to their high resistance and small volume, the dissipated energy per volume is significant even when only a small current is flowing through them. Due to poor coupling between the lattice and the electrons at low temperatures, $T_e$ can be many times the temperature of the lattice and the mixing chamber of the cryostat $T_{\mathrm{MC}}$.\cite{Roukes85} This will produce increased Johnson-Nyquist noise which is applied directly to the junction array.

\begin{figure}[t]
  \centering
  \includegraphics[width=\columnwidth]{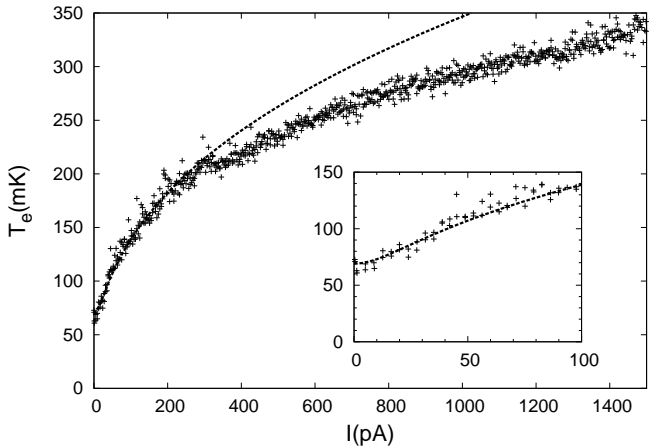}
  \caption{Effective resistor temperature vs. heating current. The inset shows a close-up of the region below 100\,pA with the axis identical to the main plot. The dashed line shows the fitted function $T_e=(\mathrm{69\,mK^5+5\times 10^6\frac{mK^5}{pA^2}}I^2)^{1/5}$, which is later used  for simulations. The fit uses only data below 300\,pA.}
  \label{fig:I_vs_T}
 \end{figure}

We probed $T_e$ in the biasing resistors using a Superconductor-Insulator-Normal metal (SIN) junction, where the superconducting electrode was made of aluminium and the normal electrode was the resistor itself. This has the advantage of probing the normal metal directly and the result is only weakly dependent on the temperatures in the superconductor.\cite{nahum94}
We calibrated this SIN thermometer against the temperature of the MC at zero current through the resistor. 
Comparing the calibration $IV$ curves with those obtained when the MC was kept at base temperature, but current was passed through the resistor, we obtain the $T_e(I)$ curve shown in fig.\,\ref{fig:I_vs_T}.
This data shows that to keep resistor temperature below 100\,mK, we need to target currents below $\approx50$\,pA. This will result in a ratio $k_BT/eV_c\approx 1/50$, which should yield reasonable noise immunity. For later use in the simulations, we have fitted the experimental data to the 5th power dependence for electron-phonon relaxation,\cite{Roukes85,wellstood94} which gives a reasonable fit when we limit it to values below $\approx 200\,\rm{mK}$. 

\section{Simulation Method}\label{sec:sim_method}
\begin{figure}[t]
 \centering
 \includegraphics[scale=0.4]{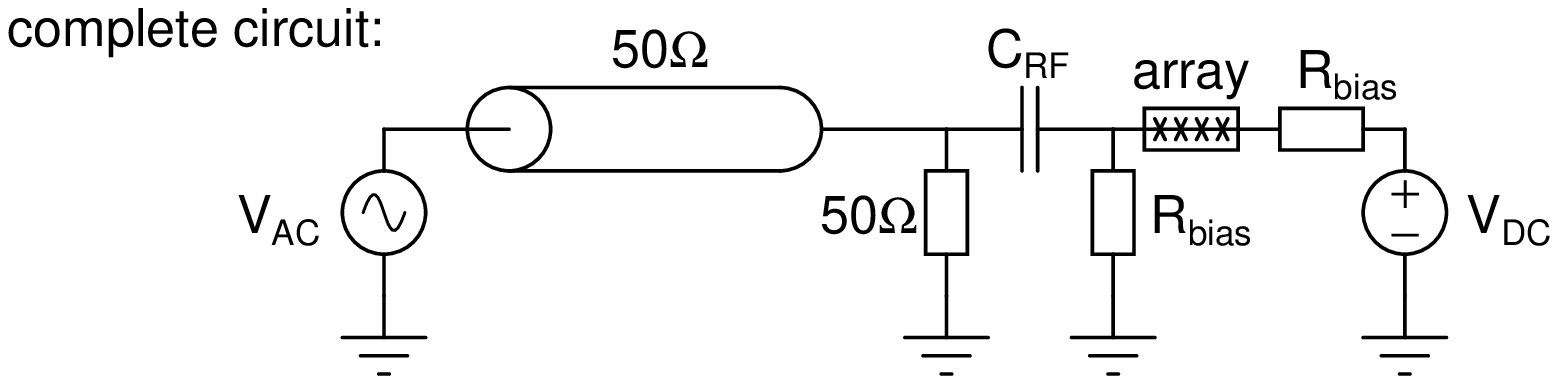}
 \includegraphics[scale=0.4]{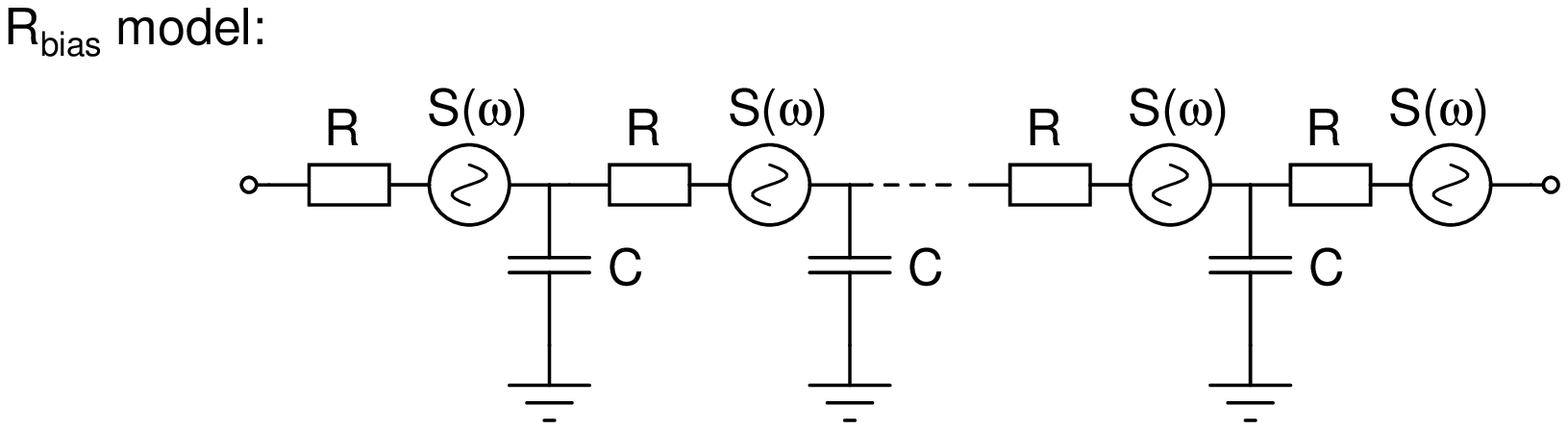}
 \includegraphics[scale=0.4]{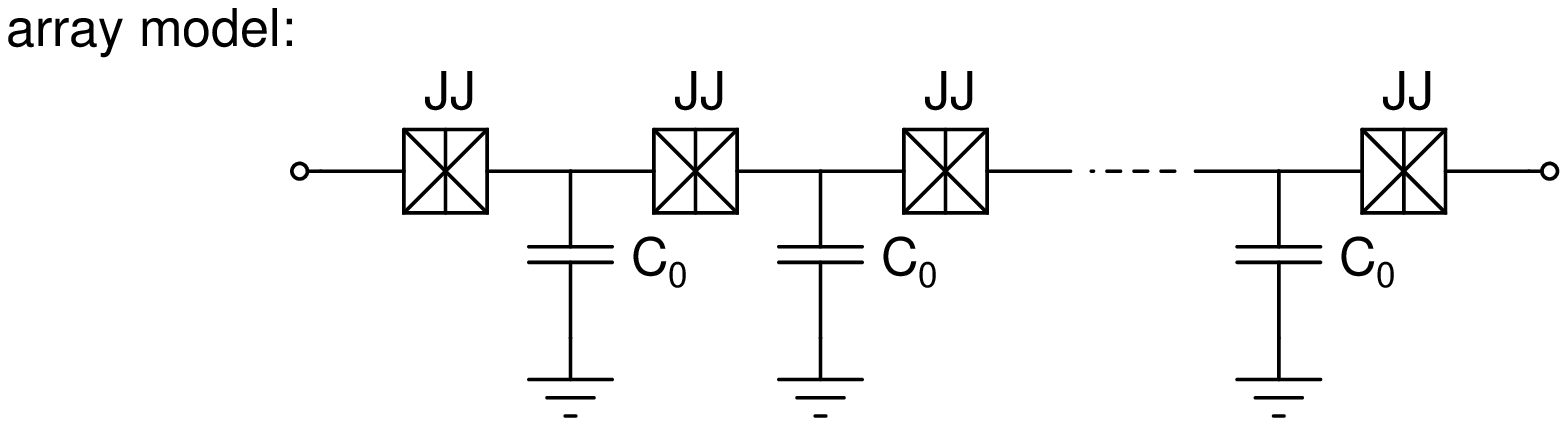}
 \includegraphics[scale=0.4]{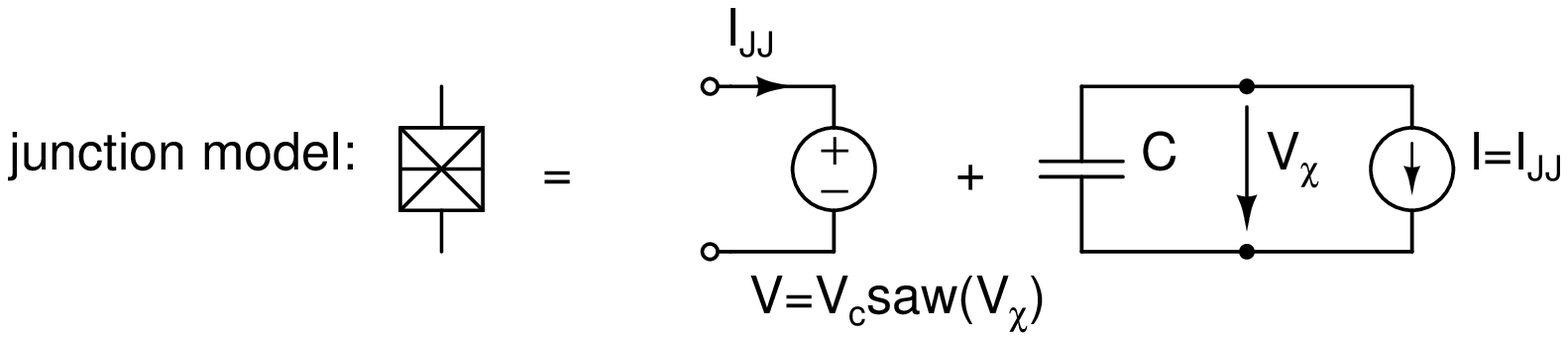}
 \caption{Circuit used for simulations including models for the noisy resistors and junction array. DC bias is applied directly, while RF is applied through the coupling capacitor $C_{RF}$.}
 \label{fig:JJB_schematic_merged}
\end{figure}

We have used the free circuit simulator ngspice\,\cite{ngspice} to model the circuit and solve eq.\,\ref{eq-motion-RC-lineSJ} in the time domain. Using a standard simulator has the advantage that arbitrary electromagnetic environments can be easily included in the simulation. Each junction is modeled as a VCVS (voltage controlled voltage source) with a periodic dependence on the normalized charge $\chi=q\frac{\pi}{e}$. This dependence can be approximated by the analytical formula
\begin{equation}
\frac{V(\chi)}{e/C} =\frac{\partial}{\partial\chi}\left[\frac{2}{\pi}\arcsin^2\left(\sqrt{\frac{1-\cos\chi}{0.3 (E_J/E_c)^2+2}}\right)\right], \label{eq:saw}
\end{equation}
applicable in the limit $0.01<E_J/E_c<1$. \cite{Agren2002} 
Since the simulator does not readily allow access to the charge as a variable, we implemented a subcircuit which integrates current through the junction into the auxiliary voltage $V_\chi$, which is then used as input voltage for the VCVS as shown in fig.\,\ref{fig:JJB_schematic_merged}.

Our model obviously excludes quasiparticle tunneling effects as well as any effects due to the upper energy bands in the junction, and thus will not show the ``Bloch nose`` exactly as seen in experiment, but it is still applicable as long as the zero-band approximation holds. The high-ohmic bias resistor is modeled as a discretized $RC$ line, where every resistor segment includes a noise source to model thermal noise as shown in fig.\,\ref{fig:JJB_schematic_merged}.

Each individual noise source is modeled in the time domain as random voltage with standard normal distribution scaled by $\sqrt{4k_BTR/\tau_{sim}}$ where $\tau_{sim}$ is the simulation time step, resulting in a white power spectrum up to the simulation bandwidth.  For the temperature of the biasing resistors the data from section\,\ref{sec:measurements} is used, and while dissipation is considered per segment, it is assumed that the resistors have a uniform temperature. Simulator bandwidth and the number of $RC$ line segments were determined individually for different simulations so that a further increase would not result in a significant change in simulator output. In practice, this would usually be the case with around 10-20\,GHz and 20 segments, respectively. Thus we fully account for the frequency dependence of the noise seen by the array for the frequencies of interest. 

\begin{figure}[t]
 \centering
 \includegraphics[width=\columnwidth]{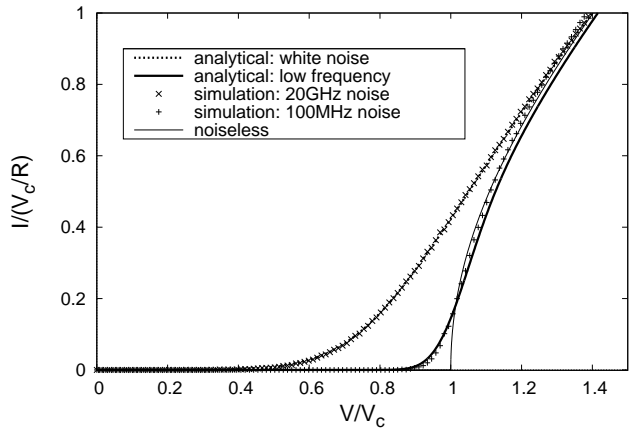}
 \caption{Comparison of the rounding of the bloackade corner by thermal fluctuations as predicted by our simulator and canonical theory for Josephson junctions in the case of white noise and low-frequency noise.}
 \label{fig:sim_vs_analytical}
\end{figure}

Correctness of the time-domain noise simulation was checked by comparing the simulator results with the canonical theory for Josephson junctions for the cases of true white noise \cite{Ivanchenko68} and low-frequency noise.\cite{kanter70} For this we used a simplified circuit which only had a single junction with almost sinusoidal $V(q)$ dependence, $V_c=761$\,$\mu$V, biased through 2\,M$\Omega$ resistance at $T=100$\,mK and zero parasitic capacitances to ground. The results are shown in fig.\,\ref{fig:sim_vs_analytical}. 
Since the time-domain simulation is inherently bandwidth limited, white noise can only be approximated by assuming a noise bandwidth significantly higher than the system bandwidth. The white noise curves match the result from theory perfectly once we take into account the fact that the noise intensity $\gamma=k_BT/E_J$, as defined in ref.\,\onlinecite{likharev-book} for the noise spectral density $S(f)=2k_BT/R$, $-\infty<f<\infty$, should be modified to $\gamma_{Bloch}=\pi k_BT/eV_c$ in our case, where the factor of $\pi$ results from the use of the normalized charge $\chi$. The low-frequency results do not require modification since the authors in  ref.\,\onlinecite{kanter70} use $I_{noise}/I_c$ as the measure of noise intensity, which we can directly transfer to $V_{noise}/V_c$ in our case. The low-frequency curves differ slightly in shape, indicating that with a noise bandwidth of 100\,MHz we are not perfectly in the low-frequency limit for our parameters yet, but the agreement is still reasonable overall.


\section{Requirements for constant-current steps}\label{sec:drive}
Dual to large junctions, where the maximum constant-voltage step for sinusoidal drive and constant damping is achieved around the characteristic frequency $f_c=I_cR_n/\Phi_0$, our arrays of ''Bloch junctions`` would exhibit the largest step around $f_c=V_c/(2eR)$, where $V_c$ is the maximum blockade voltage and $R$ is the series resistance, as long as the parasitic capacitances are negligible up to this frequency. Below $f_c$, the step size decreases similarly to the behavior of large junctions in the Josephson regime.\cite{likharev-book} This presents a problem since we want $V_c$ to be as large as possible while currents are limited to $\approx$50\,pA, resulting in a theoretical requirement for 10\,M$\Omega$ resistors. However, with our current technology it is not possible to provide such a high (real) impedance up to high frequencies due to the stray capacitance of the resistor. The useful length of the resistor is of course frequency dependent and can be roughly estimated as the length where the capacitive component becomes comparable to the resistive. For the technology values given in section\,\ref{sec:measurements}, resistors longer than tens of $\mu$m and a few hundred k$\Omega$ give diminishing returns in lowering $f_c$ and mainly serve as isolation from the warm environment.\cite{Lotkhov11}

It seems unlikely that this problem can be easily solved within the confines of current manufacturing technology. One would either need to increase the allowed DC current to $\approx 2ef_c$, or effectively lower $f_c$ further, or a combination of both. The former would require an order of magnitude increase in $E_J$ and in the efficiency of thermalization of the resistors, while the latter would require resistors with a much higher resistance per unit length and a corresponding improvement in thermalization due to the increased power dissipated per volume.
This problem arises largely due to the intrinsic coupling of the characteristic frequency of the junction (or array) and the driving frequency and consequently the resulting DC current. However, in the world of Josephson voltage standards there exists a well known method to decouple those two parameters, and that is the use of a pulse drive. In the Josephson case, this allows large steps to be achieved at pulse repetition frequencies far below $f_c$.\cite{maggi96,Benz96} The requirement for these pulses is that their rise-fall times as well as the pulse duration are about $1/f_c$ and that they have sufficient amplitude (of the order of the critical current or $\approx V_c$ for our case). As such they can be viewed as the extreme version of a sinusoidal signal at $f_c$ which is switched on for only a single half-period. In the analogy with the tilted washboard potential, this corresponds to a sudden increase in the slope of the potential, allowing the particle to move to the adjacent well, and then quickly tilting the washboard back to stop the particle from moving further. Simulation results in section\,\ref{sec:sim_results} will show that this may indeed be the method of choice to see constant-current steps with current manufacturing technology.

%
%
%

\section{Results}\label{sec:sim_results}
\begin{figure}[t]
 \centering
 \includegraphics[width=\columnwidth]{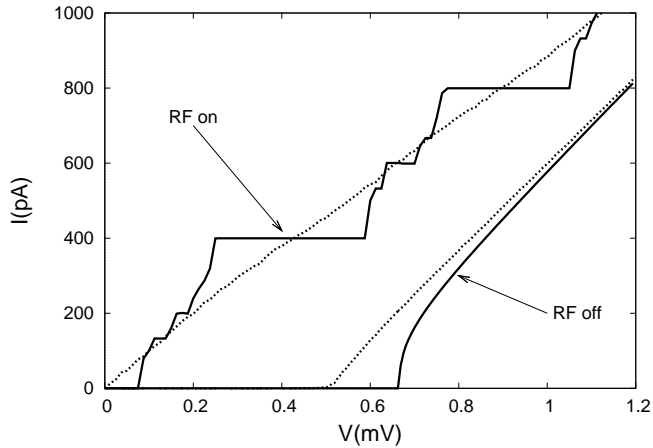}
 \caption{Simulated $IV$ curves at zero temperature (solid lines) and considering heating from the DC current (dashed lines), both with and without sinusoidal AC drive of 1.25\,GHz applied. A large step of the order of $V_c$ exists at zero temperature, but only a slight change in slope remains once realistic heating is assumed, although the blockade remains large, even in the presence of noise. The small fractional steps appear due to both the non-harmonic shape of $V(q)$ (eq.\,\ref{eq:saw}) and frequency-dependent damping (eq.\,\ref{eq-motion-RC-lineSJ}).}
 \label{fig:sim_125GHz_DC}
\end{figure}

\begin{figure}[t]
 \centering
 \includegraphics[width=\columnwidth]{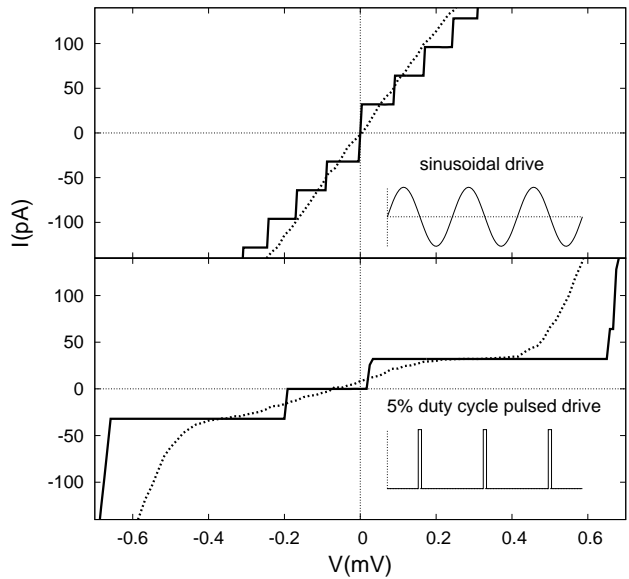}
 \caption{Comparison of simulated $IV$ curves for 100\,MHz sinusoidal drive (top panel) and pulsed drive (bottom panel). Solid lines are noiseless, dashed lines include thermal noise from the resistors. The fundamental step with sinusoidal drive at this frequency is so small that it completely disappears in the noise, while a flat step with a size of the order of $V_c$ remains with pulsed drive with 5\% duty cycle.}
 \label{fig:sim_100MHz}
\end{figure}

The different factors outlined in the previous sections conspire to make the direct observation of robust phase locking unlikely unless all factors are carefully considered in the design. In fact, we were not able to identify any set of realistic parameters where a simple sinusoidal drive leads to the observation of flat constant-current steps. The problem is one either has to run the drive at or around $f_c$, which yields a step with a width on the order of $V_c$ when zero temperature is assumed, but heats up the resistors so much that the step completely disappears once a realistic temperature is taken into account, as seen in fig.\,\ref{fig:sim_125GHz_DC}.
With our current technology the characteristic frequency $f_c$ cannot be lowered further than about 1\,GHz while retaining a reasonably large $V_c$, since the useful length of the resistors is limited by their capacitance to ground. However, just the DC current at this frequency $I=1$\,GHz$\cdot 2e\approx 320$\,nA would heat the resistors to $>200$\,mK, and including AC dissipation, this then becomes about 350\,mK.
Alternatively, one would need to run the drive significantly below $f_c$ to reduce heating, in which case the steps are much narrower even at zero temperature, just as predicted by the standard RSJ model in the Josephson case. They become so small that even though heating is significantly reduced, they again completely disappear in the noise once the temperature is taken into account, as seen in the top panel of fig.\,\ref{fig:sim_100MHz}.

A possible solution to this is to employ a pulsed drive with a low duty cycle. This effectively decouples the choice of the DC current at which to operate, from the choice of $f_c$, enabling us to fulfill the requirement for both low heating and a wide fundamental step. This is illustrated in the bottom panel of fig.\,\ref{fig:sim_100MHz}, where a pulse drive with 100\,MHz repetition rate and a 5\% duty cycle was used. In this case, the simulation yields a reasonably flat step even including heating. The AC power dissipated in the resistors in this case is in fact almost negligible due to the low duty cycle of the driving signal and barely influences the shape of the step. The pulses were simulated with rise-fall times of the simulation timestep and the duty cycle was chosen so that further decrease did not improve the step width. The precise shape of the pulses matters relatively little, as long as they are sharp and short enough, i.e. of the order of $1/f_c$. This is exactly dual to the case of the pulse-driven Josephson voltage standard as described in refs.\,\onlinecite{maggi96,Benz96}, with the one exception of the frequency-dependent damping. This is automatically modeled in our simulations and the effect can be seen in the bottom panel of fig.\,\ref{fig:sim_100MHz}: The step-width shown here for the noiseless case is the largest we were able to achieve for our circuit parameters, and it is only about $V_c$, whereas the ideal pulse-drive applied to the standard RSJ model as discussed in refs.\,\onlinecite{maggi96,Benz96} would result in about $2V_c$. 

\section{Conclusion}
Taking into account the effect of stray capacitance and power dissipation in the bias resistors as well as the specific shape of the $V(q)$ dependence of the array, we have identified a set of circuit parameters and shape of AC drive that should allow the observation of constant-current plateaus in the $IV$ curves of an array of small Josephson junctions when an AC signal is applied in addition to the DC bias, demonstrating the phase lock of Bloch oscillations and AC signal. Experiments to test this prediction are in preparation. While careful design is required, the needed parameters are entirely within reach of our fabrication technology.

Potential problems are the effects neglected in the model, in particular quasiparticle tunneling and background charges. The proximity of the resistors to the array should help in reducing the number of non-equilibrium quasiparticles.\cite{Lotkhov11} If this is not sufficient, bandgap-engineering within the array could be used, see, e.g. Refs.~\onlinecite{Aumentado04,Macleod09}. While this is typically employed to keep the single island of a single-Cooper-pair transistor free from quasiparticles by increasing its gap compared to the leads, the periodically modulated bandgap profile in an array should at least restrict the movement of quasiparticles in the array, and possibly prevent their tunneling altogether, but this needs to be experimentally determined.

In our experiments to date, the movement of background charges was quite infrequent and is thus easily tolerable for this application as long as the $V(q)$-dependence remains single-valued with still reasonable blockade amplitude (see fig.\,\ref{fig:Fig_V-vs-q}).

Delivering a properly shaped drive signal into a high-impedance environment at mK temperatures is - while challenging - routinely done for qubit setups.

Finally, the results of this work can also be applied to the problem of Shapiro-like steps in the IV-curves of superconducting nanowires embedded in a high-impedance environment. As has been predicted by Mooij and Nazarov,\cite{MooijNazarov} the
effect of quantum phase slips in these circuits may result in coherent motion of Cooper pairs through the wire, which could then be phase locked to an external drive, yielding current steps at $I = 2ef$. The behavior of such a nanowire
is similar to that of a short array of small Josephson junctions.\cite{Matveev} This was recently 
demonstrated in an experiment with a Josephson junction array in a ring configuration.\cite{Pop2010} Thus, an experiment aiming to demonstrate such a phase lock in nanowires will face the same challenges and the design will need to account for the issues of frequency dependent damping and overheating of the biasing resistors, which we addressed in this paper.

\section{Acknowledgements}
The authors would like to thank Oliver Kieler, Jukka Pekola, Olli-Pentti Saira, and Thomas Scheller for fruitful discussions, as well as  the ngspice developers, especially Holger Vogt and Robert Larice, for continuously improving the simulator and quickly ironing out the bugs encountered in this work. 

This work was in part financed through the SCOPE project by the Future and Emerging Technologies (FET) programme within the Seventh Framework Programme for Research of the European Commission, under FET-Open grant number 218783.

\bibliography{Maibaum_Zorin-Bloch_sim}%

\end{document}